\documentclass[aps,preprintnumbers,amsmath,amssymb,prd,longbibliography,twocolumn]{revtex4-2}

\usepackage{dcolumn}
\usepackage{color}
\usepackage{subfigure}  
\def\comment#1{}

\newcommand{\nc}{\newcommand}
\nc{\beq}{\begin{eqnarray}}
	\nc{\eeq}{\end{eqnarray}}
\nc{\scs}{\scriptstyle}
\nc{\setval}{\fmfset{wiggly_len}{3mm} \fmfset{arrow_len}{1.5mm}
	\fmfset{arrow_ang}{13} \fmfset{dash_len}{1.5mm}\fmfpen{0.125mm}
	\fmfset{dot_size}{2thick}}

\usepackage{bm,latexsym,mathrsfs,enumerate,color}
\usepackage[mathcal]{euscript}
\usepackage[breaklinks=true,unicode=true,urlcolor = blue,colorlinks = true,citecolor = blue,linkcolor = blue]{hyperref}
\usepackage{graphicx}
\usepackage{wrapfig}
\usepackage{dsfont}
\usepackage{bbm}
\usepackage[percent]{overpic}
\usepackage{empheq}

\usepackage{outlines}
\usepackage[capitalize]{cleveref}
\renewcommand{\vec}[1]{\bm{#1}}

\def\slashchar#1{\setbox0=\hbox{$#1$}           
	\dimen0=\wd0                                 
	\setbox1=\hbox{/} \dimen1=\wd1               
	\ifdim\dimen0>\dimen1                        
	\rlap{\hbox to \dimen0{\hfil/\hfil}}      
	#1                                        
	\else                                        
	\rlap{\hbox to \dimen1{\hfil$#1$\hfil}}   
	/                                         
	\fi}                                         %

\DeclareMathAlphabet\mathbfcal{OMS}{cmsy}{b}{n}

\usepackage{mathtools}
\usepackage{graphicx}
\usepackage{dcolumn}
\usepackage{amssymb}
\usepackage{amsmath}
\usepackage{slashed}
\usepackage{bm}
\usepackage{bbm}

\begin{document}

\title{Multi-critical point and unified description of broken-symmetry phases in spin-$\frac{1}{2}$ anti-ferromagnets on a square lattice}
\author{O\u{g}uz T\"urker} 
\author{ Kun Yang}
\affiliation{National High Magnetic Field Laboratory and Department of Physics,
Florida State University, Tallahassee, Florida 32306, USA}

\begin{abstract}

We show that several distinct broken-symmetry phases in a spin-$\frac{1}{2}$ anti-ferromagnet on a square lattice with easy-plane anisotropy, including valence bond solid, chiral spin liquid, and the XY-ordered state, can all be accessed by perturbing a multi-critical point with two massless Dirac fermions coupled to a level-one Chern-Simons gauge field. This allows for a unified description of these phases, as well as the phase transitions between them. In a specific phase transition, our analysis provides a lattice realization of one of the recently proposed fermion-boson dualities, thus lending support to it. We also briefly discuss the relation between our paper and the long-sought deconfined criticality in such systems.

\end{abstract}

\date{\today}

\maketitle

\section{Introduction}

Two-dimensional (2D) spin-$\frac{1}{2}$ antiferromagnets can support a large variety of phases, many of them break spin-rotation and/or lattice symmetries. Spin liquids\cite{KivelsonSpinLiquidReview}, which break none of these symmetries, have been the focus of much recent research activity. They come in many different types as well. One such type, known as chiral spin liquid that breaks time-reversal symmetry, will be of particular relevance to our discussion below. Needless to say, quantum phase transitions among all these phases are also of strong interest.

Broken symmetry phases are traditionally described in Ginzburg-Landau theory, which is a field theory written in terms of the local order parameter associated with the spontaneously broken symmetry (see, { e.g.}, Ref. \cite{BOOK}). In such descriptions, phases with different broken symmetries are described using different order parameter fields, and direct second-order transitions between them require fine-tuning. Instead, the more generic situations are first-order transitions or intermediate phases where both types of orders co-exist. In Ref. \cite{deconfined_cirticalityPRB}, Senthil and \emph{et.al.} argue that such descriptions miss the possibility of deconfined criticality, which is a critical point separating two different broken symmetries facilitating a direct second-order transition between them. Such novel quantum criticality can only be captured in a field theory that describe both types of broken symmetries on equal footing. Specifically, they argue that such deconfined critical points separate the Neel ordered and valence bond solid (VBS) phases of 2D spin-$\frac{1}{2}$ antiferromagnets, which break spin-rotation and lattice translation symmetry, respectively. In the appropriate field theory, the two symmetry-breaking order parameters are dual to each other and thus afford a unified description. Numerous attempts have been made to identify such deconfined critical points, with inconclusive outcomes thus far (see Refs. \cite{PhysRevLett.125.257204} and \cite{PhysRevB.102.195135} for recent attempts for the Heisenberg and XY symmetry classes, respectively, and references therein).

While it is {\em not} our goal to resolve the fate of deconfined criticality, our work is motivated by the line of thoughts that lead to it. To this end, we seek to find a field theory that provides a unified description of relevant phases in this description, and beyond. We find by perturbing a theory of two massless Dirac fermions coupled to a single level-one Chern-Simons (CS) gauge field, we can reach the XY-ordered (we only consider 2D spin-$\frac{1}{2}$ antiferromagnets with easy-plane anisotropy in this paper), VBS, chiral spin liquid, and an Ising Neel state in which the Neel order is along the $z$-direction despite the easy-plane anisotropy (which is possible in the presence of frustration). Within this description, a direct second-order transition between the XY-ordered phase and the VBS or Ising Neel phase must go through this massless point, which requires fine-tuning.

The remainder of the paper is organized as  follows. In Sec. \ref{sec:model}, we introduce the spin-$\frac{1}{2}$ XY model and arrive at the multi-critical point by attaching a flux quantum to each hard-core boson that represents an up spin, and perform a mean-field approximation to smear out the flux. This results in two massless Dirac fermions coupled to a level-one CS field. In Sec. \ref{sec:Dirac masses}, we discuss the phases that result when various mass terms are added to perturb this critical point. In Sec. \ref{sec:ssb}, we discuss how the mass terms responsible for spontaneous lattice symmetry breaking are generated by fermion interactions. Section \ref{sec:Dual Description} is devoted to deriving the dual bosonic theory of the multi-critical point, where we also make comparison with the existing theory of de-confined criticality. A brief summary is offered in Sec. \ref{sec:Summary}.

\section{Model and Composite Fermion Mean-field Approximation}
\label{sec:model}

We consider the following spin-$\frac{1}{2}$ Hamiltonian on the square lattice:
\begin{align}
H&=-\sum_{<ij>}(S_i^x S_j^x + S_i^y S_j^y) + \cdots \label{eq:model}\\
 &= H_0 + \cdots\label{eq:model1},
\end{align}
where $<ij>$ stands for nearest-neighbors, and the ellipsis represents generic additional couplings that respect the XY rotation symmetry and all lattice symmetries unless noted otherwise. Note that the minus sign means the XY coupling is ferromagnetic instead of anti-ferromagnetic; the two are equivalent under a $\pi$ rotation along the $z$-direction for one of the two sub-lattices. The advantage of considering the ferromagnetic XY coupling is the XY-ordered phase only breaks the $O(2)$ spin rotation symmetry, but none of the lattice symmetries. This makes the discussion of broken symmetries in various phases simpler below. The antiferromagnetic nature of \cref{eq:model} is thus hidden in the ellipsis, which include $S^z$ and further neighbor couplings between XY spins in the same sublattice.

We can map half-spin ladder operators to annihilation and creation operators  of the  hard-core bosons. Accordingly, the nearest-neighbor XY spin coupling in \cref{eq:model} becomes nearest-neighbor boson hopping,
\begin{eqnarray}
\label{eq:model2}
H_0=-\frac{1}{2}\sum_{<ij>}(b^\dagger_i b_j + b^\dagger_j b_i),
\end{eqnarray}
and the ground state has half-filling in the absence of a net magnetization along the $z$-direction. We will use the spin and boson representations interchangeably below.

\begin{figure}
\centering
\includegraphics{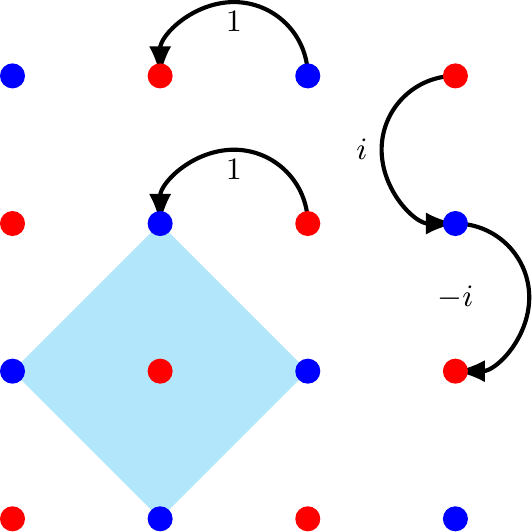}

\caption{Gauge choice for fermion hopping phases. The magnetic unit cell contains two squares, two lattice sites (one each from A and B sublattices), and a total of $2\pi$ flux. For bonds with imaginary phases, the phase corresponds to hopping in the direction of the error.}
\label{fig:1}
\end{figure}

To proceed, we map the hard-core bosons to composite fermions (CFs) attached to a flux quantum by coupling them with pure CS theory in lattice, and then make a mean-field approximation to spread out the flux uniformly that results in a $\pi$ (or half) flux per plaquette \cite{Fradkin94}. With the gauge choice of Fig. \ref{fig:1}, the resultant band Hamiltonian takes the form
\begin{equation}
h_{\vec{k}} = \left(\begin{array}{cc}
   0 & \sin k_x + i\sin k_y\\
   \sin k_x - i\sin k_y & 0\end{array}\right),
\label{eq:MF Hamiltonian}
\end{equation}
in which $\vec{k}$ is the lattice momentum. Importantly, we have two Dirac points at $(0, 0)$ and $(\pi, 0)$ where the two bands meet. In the ground state, the lower band is filled while the upper band is empty, so the chemical potential coincides with the Dirac points. Thus, the low-energy physics of the system at this level of approximation is described by two species of massless Dirac fermions coupled to a {\em single}CS gauge field:
\begin{equation}
\mathcal{L} = i\bar{\Psi}\slashed{D}\Psi + \mathcal{L}_\text{CS}[a] +\cdots,
\label{eq:Low-energy theory}
\end{equation}
where,
\begin{equation}
\mathcal{L}_\text{CS}[a] =  \frac{1}{4\pi} \epsilon^{\mu\nu\lambda}a_\mu\partial_\nu a_\lambda = a\wedge da,
\label{eq:LCS term}
\end{equation}
is the level-one CS term, $\Psi = (\psi_1, \psi_2)^T$ combines the two Dirac fields \footnote{One for each Dirac point, and they each have two components representing the A and B sublattice. Note a rotation is performed on one of the Dirac points so that the two Dirac fields have the same chirality which allows for a unified description in \cref{eq:Low-energy theory}.} where $\psi_i$ are two-component Dirac spinors, the slash notation is defined for a general three-vector $b_\mu$ as $\slashed{b}=\gamma^\mu b_\mu$, where $\gamma^\mu$ are two by two Dirac matrices obeying the Clifford algebra $\{\gamma^\mu,\gamma^\nu\}=2\eta^{\mu\nu}$, where $\eta^{\mu\nu}$ is the metric of the Minkowski $2+1$ space-time and $\eta^{\mu\nu}=\text{diag}(+,-,-)$, which will be used for raising and lowering the indices throughout the paper and $\{,\}$ is the anti-commutator. $D_\mu = \partial_\mu -ia_\mu -iA_\mu$ includes  coupling to both the dynamic field $a_\mu$ and background gauge field $A_\mu$, and the ellipsis represents the less relevant terms like the Maxwell term of $a$. Equation (\ref{eq:Low-energy theory}) is the same theory discussed in Ref. \onlinecite{BarkeshliMcGreevy} in a closely related context. As we demonstrate below, a variety of interesting phases supported by \cref{eq:model} can be accessed by perturbing \cref{eq:MF Hamiltonian} with various mass terms for the Dirac fermions.

\section{Dirac Mass Terms and Corresponding Broken Symmetry Phases}
\label{sec:Dirac masses}

The most general mass term that couples the two Dirac points takes the form $\bar{\Psi}M\Psi$, where
\begin{equation}
M = m_0 \mathbbm{1} + m_1 \sigma_1 + m_2 \sigma_2 + m_3 \sigma_3 = m_0\mathbbm{1}  + \vec{ m}\cdot\vec{ \sigma}
\label{eq:Dirac Mass}
\end{equation}
is a two-by-two Hermitian matrix. In the following, we discuss how such mass terms can be generated beyond the mean-field Hamiltonian \cref{eq:MF Hamiltonian}, and what phases they generate once added to the critical theory \cref{eq:Low-energy theory}.

\subsection{Uniform Mass $m_0$}
\label{subsec: m0}

We first consider next-nearest-neighbor XY coupling,
\begin{eqnarray}
\label{eq:nnnCoupling}
H_{\text{nnn}}&=&J'\sum_{<<ij>>}(S_i^x S_j^x + S_i^y S_j^y)\\
& =& J'\sum_{<<ij>>}(b^\dagger_i b_j + b^\dagger_j b_i),
\end{eqnarray}
where $<<ij>>$ stands for next-nearest-neighbors. Within the mean-field approximation and using the gauge choice that hopping between A sublattice sites have phase +1 and that between B sublattice sites have phase $-$1   results in a term of the form
\begin{equation}
h'_{\vec{k}} = 2J'\left(\begin{array}{cc}
   \cos k_x \cos k_y & 0\\
   0 & - \cos k_x \cos k_y\end{array}\right),
\label{eq:nnnCouplingFermion}
\end{equation}
resulting in a uniform mass term with $m_0 = 2J'$, while $\vec{m} =0$.

We now analyze the phases stabilized by $m_0\ne 0$. Since the fermions are massive, they can be integrated out. This results in a CS term $\text{sgn} (m_0)\mathcal{L}_\text{CS}[A+a]$ whose sign depends on that of $m_0$ or, equivalently, $J'$, which needs to be combined with the original CS term for $a$ in \cref{eq:Low-energy theory}. We analyze the two cases separately.

$\bullet$(i) $m_0 < 0$. In this case, we have
\begin{equation}
\mathcal{L}_\text{eff}[a, A] = \mathcal{L}_\text{CS}[a] - \mathcal{L}_\text{CS}[a+A] = - 2 a\wedge dA - \mathcal{L}_\text{CS}[A].
\end{equation}
Since the CS coupling of $a$ gets canceled, we are left with a linear coupling between $a$ and $A$. Further integrating out $a$ yields a constraint $dA=0$. This corresponds to a Meissner response of the hard core bosons, indicating they are in a superfluid phase that spontaneously breaks the U(1) symmetry that corresponds to charge conservation \cite{BarkeshliMcGreevy}. For the original spin-$\frac{1}{2}$ Hamiltonian \cref{eq:model}, this is the XY-ordered phase \cite{Fradkin94}.

$\bullet$(ii) $m_0 > 0$. In this case, we have
\begin{equation}
\mathcal{L}_\text{eff}[a, A] = \mathcal{L}_\text{CS}[a] + \mathcal{L}_\text{CS}[a+A] = 2 a\wedge da + 2 a\wedge dA + \mathcal{L}_\text{CS}[A].
\end{equation}
Further integrating out $a$ yields
\begin{equation}
\mathcal{L}_\text{eff}[A] = -\frac{1}{2} A\wedge dA + \mathcal{L}_\text{CS}[A] = \frac{1}{2} \mathcal{L}_\text{CS}[A].
\end{equation}
This is a fractional quantum Hall response corresponding to the $\nu=\frac{1}{2}$ Laughlin state for bosons \cite{BarkeshliMcGreevy}. In the original spin model, this corresponds to the Kalmeyer-Laughlin chiral spin liquid (CSL) state, in which time-reversal symmetry is spontaneously broken. The same result was obtained earlier on  triangular and Kagom\'{e} lattices with anti-ferromagnetic nearest-neighbor XY coupling only \cite{Yang93}.

It should be noted that while the mean-field Hamiltonian Eq. (\ref{eq:MF Hamiltonian}) suggests that the nearest-neighbor XY model is a critical point separating the CSL and XY-ordered phases, it is known that its ground state is actually XY ordered. As discussed in Ref. [\onlinecite{Fradkin94}], fluctuation effects beyond the mean-field approximation tend to generate a negative $m_0$. We thus need a positive $J'$, which {\em frustrates} the XY order, to reduce the magnitude of the dynamically generated negative mass, and eventually drive the system into the CSL phase. It would be very interesting to study the spin-$\frac{1}{2}$ XY model with the frustrating next-nearest-neighbor $J'$ coupling to see if such a transition exists.

\subsection{Staggered Mass $m_3$}

We call $m_3$ in Eq. (\ref{eq:Dirac Mass}) staggered mass because it gives rise to masses of opposite sign to the two Dirac fermions. Interestingly, it comes from a staggered potential coupled to the hardcore boson density (which is equal to the CF density),
\begin{eqnarray}
\label{eq:staggered mass}
v=m_3\sum_i (-1)^i b^\dagger_i b_i = 2m_3\sum_i (-1)^i S^z_i + \text{const},
\end{eqnarray}
and the second equality above indicates it couples to the staggered magnetization along $z$ direction in the original spin language. Such a mass could come from spontaneous development of staggered magnetization along the  $z$ direction \cite{Fradkin94}, which breaks lattice translation symmetry {\em spontaneously}. We call the resultant phase an Ising-ordered phase (to be distinguished from the XY-ordered phase discussed earlier). $m_3$ could also come from an external potential with wave vector $(\pi, \pi)$, which breaks lattice translation symmetry {\em explicitly}.

Regardless of its origin, in the presence of $m_3$ the Dirac fermions can again be integrated out. Since they have opposite masses, the CS terms they generate cancel. Further integrating out $a$ with the existing CS term thus generates no term involving $A$, indicating the state has no (non-trivial) electromagnetic response. This is thus a Mott insulator state for the hardcore bosons.

Reference [\onlinecite{Fradkin94}] was mainly concerned about the quantum phase transition from the XY-ordered to Ising-ordered state in the nearest-neighbor XXZ model, which is actually a first-order transition that occurs at the Heisenberg point. In the presence of frustration, like that induced by $J'$, XY order gets suppressed and a direct second-order transition between them may be possible. Since $m_3$ breaks lattice translation symmetry, it must remain zero at this (putative) critical point. As a result, the transition must again be driven through the critical point described by Eq. (\ref{eq:Low-energy theory}), where $m_0$ vanishes and $m_3$ gets turned on simultaneously. This is different from the conclusion of Ref.[\onlinecite{Fradkin94}], where the authors assumed the presence of both $m_0$ and $m_3$, resulting in masses $m_0 \pm m_3$ for the two Dirac fermions, and the critical point is reached at $m_0 = m_3$, where only one of the two Dirac fermions become massless,
\begin{equation}
\mathcal{L} = i\bar{\psi}\slashed{D}\psi + \mathcal{L}_\text{CS}[a] +\cdots,
\label{eq:Low-energy theory with single fermion}
\end{equation}
where $\psi$ is the field of this massless Dirac fermion. This same model  was also discussed in Ref. \cite{Flavio2021}.

From the discussions above, it becomes clear that for the theory [\cref{eq:Low-energy theory with single fermion}] to be relevant, before the XY order is suppressed, the Ising order must be present already, due to either spontaneous or explicit breaking of lattice translation symmetry. We consider the latter for its simplicity. With a staggered lattice potential of the form Eq. (\ref{eq:staggered mass}), the unit cell of the square lattice gets doubled, and so does the boson filling from half to one per unit cell. We thus have a standard super-fluid (SF) to Mott-Insulator transition in this case, which is described by the familiar $O$(2) $\phi^4$ theory. Our analysis thus support the recently proposed duality between \cref{eq:Low-energy theory with single fermion} and the corresponding Wilson-Fisher fixed point \cite{DualityReview}.

\subsection{Off-diagonal Masses $m_1$ and $m_2$}

As discussed above, the staggered mass $m_3$ breaks lattice translation symmetry and carries momentum $(\pi, \pi)$. The off-diagonal mass terms $m_1$ and $m_2$ couple the two Dirac points, and must carry momentum $(\pi, 0)$ or $(0, \pi)$. They thus break lattice translation symmetry in a different manner. As we demonstrate below, column (VBS) orders correspond to such symmetry breaking pattern and generates these masses.

\begin{figure}
	\centering
	\includegraphics{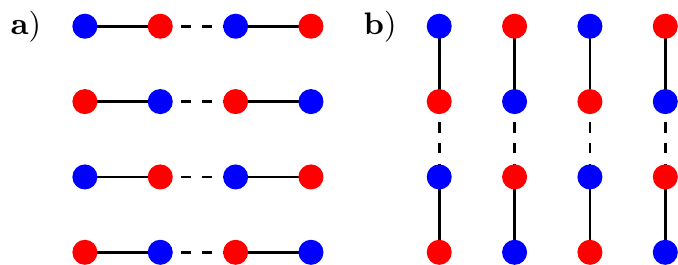}
	\caption{Column valence bond solid (VBS) patterns. Solid lines represent strengthened bonds while dashed lines represent weakened bonds. (a) The column VBS directed along $x$ axis, carrying lattice momentum $(\pi, 0)$ .(b) The column VBS, directed along $y$ axis, carrying lattice momentum $(0, \pi)$. }
	\label{fig:2}
\end{figure}

VBS order, generated either spontaneously or explicitly, modulates the spin-spin coupling strength. We consider the most important column VBS patterns, which could align along either the $x$ or $y$ direction (see Fig. \ref{fig:2}). It is immediately clear that they carry momenta $(\pi, 0)$ and $(0, \pi)$, respectively. A straightforward calculation yields $m_{2,3} = \delta$ for the patterns of Figs. \ref{fig:2}(a) and \ref{fig:2}(b) respectively, where $\delta$ is the bond modulation.

In the presence of $m_2$ and/or $m_3$, we can diagonalize the mass matrix \cref{eq:Dirac Mass} resulting in the same Dirac Hamiltonian as that of $m_3$ mass only. So the system has the same topological properties as well, and is in the Mott insulator phase.

\subsection{Summary}

\begin{figure}
	\centering
\includegraphics{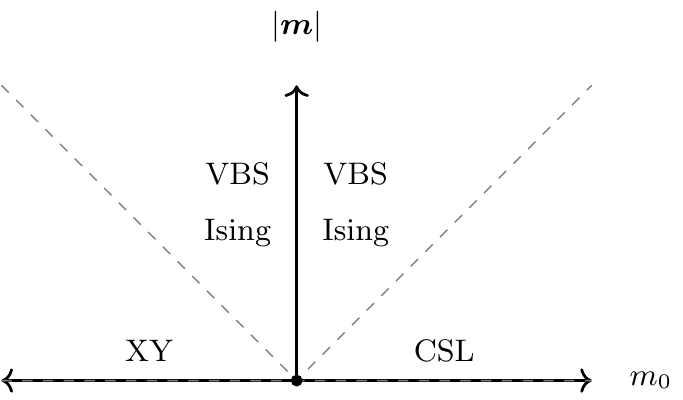}
\caption{Phase diagram parameterized by the mass matrix of \cref{eq:Dirac Mass}. The dashed lines are second-order phase boundaries along which one of the eigenvalues of the mass matrix Eq. (\ref{eq:Dirac Mass}) vanishes. The origin (thick dot) is a multi-critical point where the entire mass matrix Eq. (\ref{eq:Dirac Mass}) vanishes. See text for detailed description of phases and the symmetries they break.}
\label{fig:3}
\end{figure}

When all masses are present, diagonalizing the mass matrix Eq. (\ref{eq:Dirac Mass}) yields
\begin{equation}
m= m_0\pm\sqrt{m_1^2 + m_2^2 + m_3^2} = m_0\pm|\vec{m}|,
\end{equation}
where $\vec{m}=(m_1, m_2, m_3)$. The resultant phase diagram, Fig. \ref{fig:3}, looks similar to that of Ref. [\onlinecite{BarkeshliMcGreevy}], although what we have here is actually a 4D phase diagram projected down to the 2D plane spanned by $m_0$ and $|\vec{m}|$. In particular, we note all components of $\vec{m}$ break lattice translation symmetry, but in different ways. On the other hand, $m_0$, while not breaking lattice translation symmetry, leads to phases that breaks $O$(2) spin rotation symmetry or time-reversal symmetry when it dominates, in the same manner as discussed in Sec. \ref{subsec: m0}.

In our description, the two perpendicular VBS order parameters give rise to the real and imaginary parts of the off-diagonal Dirac mass in Eq.  (\ref{eq:Dirac Mass}). They thus naturally form a complex order parameter, consistent with an earlier study \cite{deconfined_cirticalityPRB}. We find they can be further combined with stagger magnetization $m_3$ to form an $O$(3) order parameter, and they cooperate to enhance the Mott gap; in other words, they are intertwined. On the other hand, they compete with the uniform mass $m_0$, and such competition leads to various quantum phase transitions. Despite such competition, our analysis suggests that all the phases that appear in Fig. \ref{fig:3} naturally appear near each other in a frustrated spin-$\frac{1}{2}$ model on the square lattice, in the neighborhood of a multi-critical point described by Eq. (\ref{eq:Low-energy theory}).

\section{Spontaneous Breaking of Lattice Symmetry}\label{sec:ssb}

From the perspective of the field theory Eq. (\ref{eq:Low-energy theory}), the massless point for both of the Dirac fermions (the origin in Fig. \ref{fig:3}) is multi-critical and the full mass matrix of Eq. (\ref{eq:Dirac Mass}) must be tuned to zero. For example a direct second-order transition from the XY phase to the VBS phase must go through this point, while a more generic situation is going through the co-existing region or a direct first-order transition.
On the other hand, the masses $\vec{m}$ break lattice symmetries. Thus, unlike $m_0$, they are not tuning parameters, but are instead generated from (sufficiently) strong interactions that lead to spontaneous symmetry breaking. Accordingly, we consider the following four-Fermi (Gross-Neveu type) interaction:
\begin{eqnarray}
&\mathcal{L}_{\text{int}} &= \lambda_0[(\overline{\psi}_1\psi_1)^2 + (\overline{\psi}_2{\psi}_2)^2] + \lambda_1(\overline{\psi}_1{\psi}_1)(\overline{\psi}_2{\psi}_2)\nonumber\\
 &+& \lambda_2(\overline{\psi}_1{\psi}_2)(\overline{\psi}_2{\psi}_1) + \lambda_3 (\overline{\psi}_1\psi_2)^2 + \lambda^*_3 (\overline{\psi}_2\psi_1)^2.
\label{eq:GN interaction}
\end{eqnarray}
It is clear that a positive $\lambda_1$ favours $m_3$, while a negative $\lambda_2$ and any $\lambda_3$ favour $m_{1,2}$. We can introduce Hubbard-Stratonovich fields $\Phi$ to decouple these interactions, resulting in a Yukawa type of coupling
\begin{equation}
\mathcal{L}_Y = \bar{\Psi}\Phi\Psi,
\end{equation}
where
\begin{equation}
\Phi = \phi_0 \mathbbm{1} + \phi_1 \sigma_1 + \phi_2 \sigma_2 + \phi_3 \sigma_3.
\label{eq:Yukawa matrix}
\end{equation}
Obviously $\vec{\phi} = (\phi_1, \phi_2, \phi_3)$ is an order parameter field describing the broken lattice symmetry.

The ordering transition is described by an effective field theory in terms of $\vec{\phi}$ obtained from integrating out $\Psi$.
This can be done under the generic situation of $m_0\ne 0$. Such a transition, if continuous, takes the system from the XY/CSL phase to a mixed phase where spontaneously broken XY/time-reversal symmetry coexist with spontaneously broken lattice symmetry. A direct continuous transition from the XY phase to the VBS phase, however, again requires fine-tuning $m_0$ to zero; in this case, the Dirac fermions are massless and cannot be integrated out perturbatively.

Returning to the generic situation of $m_0\ne 0$, to determine the order of transition this effective theory describes at the mean-field level, we are interested in the sign of the prefactor of the $|\vec{\phi}|^4$ term. Let us denote this prefactor as $\beta_4$. We can calculate it diagrammatically.  For notational simplicity, we focus on the $\phi_3$ term in \cref{eq:Yukawa matrix} as a representative of $\vec{\phi}$; the conclusions below are general. Also, we assume a uniform $|\vec{\phi}|=m$.

\begin{figure}[ht!]	
	
\includegraphics{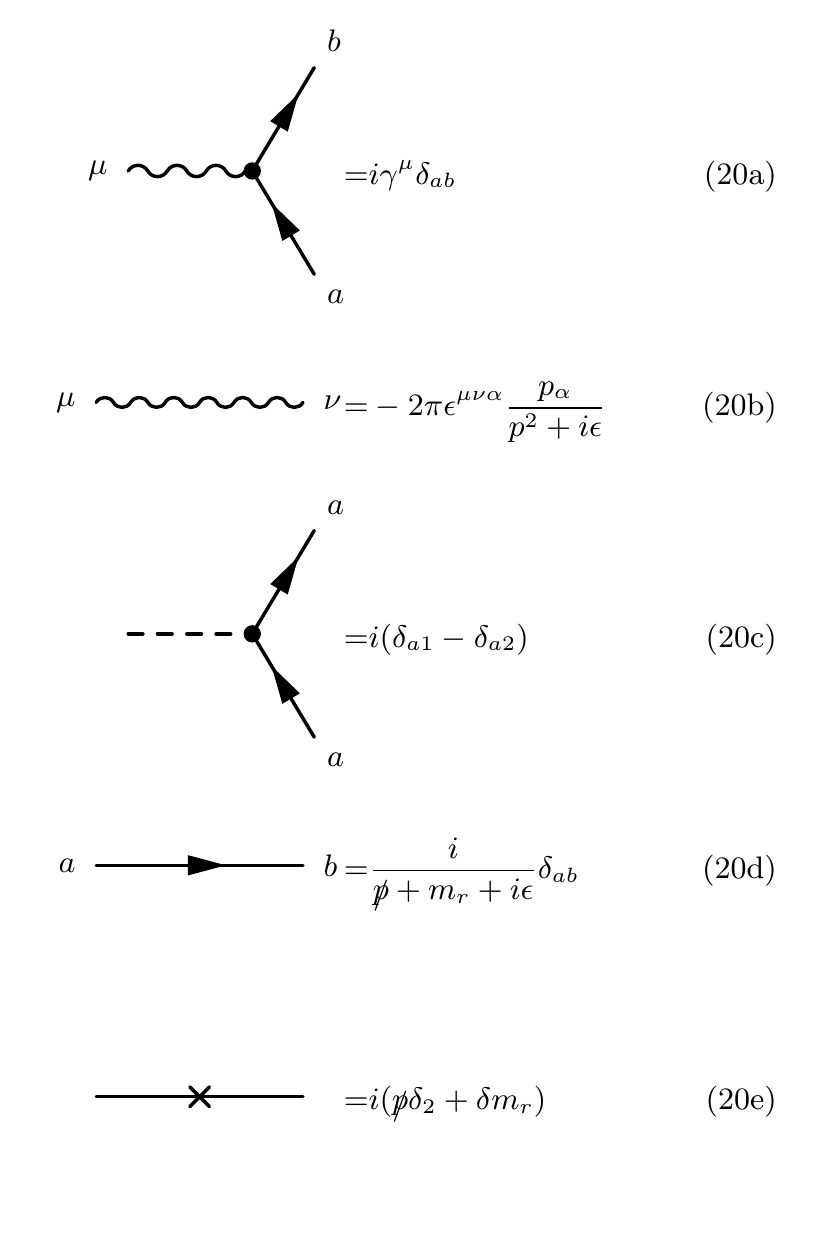}

	\caption{Feynman rules of our model, where $a,b$ denote fermion flavors. Dashed lines represent  $\phi_3$, solid lines represent fermions, and curly lines represent the gauge field $a_\mu$ (we use Landau gauge), and finally the cross is the counter term which we denote as $D_5$, where $\delta_2$ is the fermion field strength renormalization counter-term and  $\delta_m$ is the mass counter-term (see also Fig. \ref{fig:els}). We will drop the arrows in the fermionic lines in the rest of the paper, to have less cluttered diagrams.}\label{fig:feynmanrules}
\end{figure}

\begin{figure}[ht!]
\includegraphics{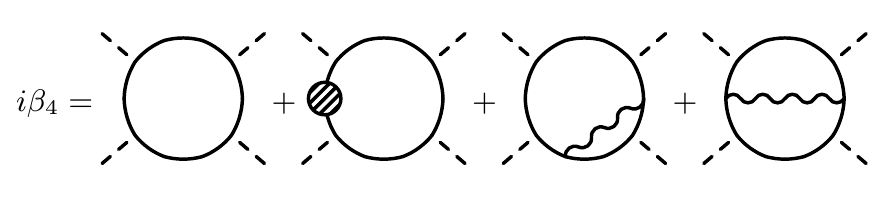}
	\caption{Diagrammatic representation of the $\beta_4$ up to two loops. We denote these diagrams, respectively, from left to right as, $D_1$, $D_2$, $D_3$, $D_4$ and the shaded blob is the self-energy which is shown in \cref{fig:els}.}\label{fig:phi4}
\end{figure}

The Feynman rules we use are given in \cref{fig:feynmanrules}.
The diagrams up to two loops that contribute to $\beta_4$ are shown in \cref{fig:phi4}\footnote{There are other possible four external leg and two-loop, irreducible diagrams with one photon propagator. We find those diagrams to be zero under an appropriate regularization scheme.}.
The expansion in the number of loops is equivalent to a weak coupling expansion in terms of the coupling constant between fermions and CS gauge field \cite{Chen1993}, the inverse of the square root of the absolute value of the CS level. We do not show this coupling constant explicitly in our calculations for brevity of the notation.
We adopt a renormalized perturbation theory approach, in which we replace the bare mass $m_0$ with renormalized mass $m_r$ in the free propagators and compensate this by adding mass and field-strength renormalization counter-terms.

Before the  quantitative  computation of $\beta_4$, we can have a qualitative discussion on what to expect. The dimension of the $\beta_4$ has an inverse mass dimension, i.e., $[\beta_4]=[m^{-1}]$. On the other hand, the only dimensionful free parameter in the theory (and relevant Feynman diagrams that generate $\beta_4$) is the mass $m_r$. We thus expect $\beta_4$ is inversely proportional to the mass. We will calculate this proportionality constant below.

We start with calculating the one-loop contribution $D_1$ which is shown in \cref{fig:phi4}:
\begin{align}
	D_1&=-\frac{1}{4}\text{tr} \Bigg[\int \frac{d^3p}{(2\pi)^3}\frac{(i(\slashed{p}-m_r))^4}{(p^2-m_r^2)^4+i\epsilon}\Bigg]{\notag}\\
	&=-\frac{1}{4}\text{tr}\Bigg[\int\frac{d^3p}{(2\pi)^3}\frac{p^4-4p^2m_{r}\slashed{p}+6p^2m^2-4\slashed{p}m_{r}^3+m_{r}^4}{(p^2-m_r^2)^4+i\epsilon}\Bigg]
	{\notag}\\
	&=-\frac{2}{4}\int\frac{d^3p}{(2\pi)^3}\frac{p^4+6p^2m_{r}^2+m_{r}^4}{(p^2-m_r^2)^4+i\epsilon},
\end{align}
which, after Wick rotating to Euclidean coordinates $p^{\mu}_E=p_{E \mu }=(-ip^0,\vec{p})$ and $\slashed{p}=-i\slashed{p}_E$, becomes
\begin{align}
	D_1&=-\frac{i}{2}\frac{1}{|m_r|}\int\frac{d^3p_E}{(2\pi)^3}\frac{p_E^4-6p_E^2+1}{({(p_E)}^2+1)^4}.
\end{align}
Using the integral identities given in \cref{sec:inti} with $n=4$, $d=3$ we find,
\begin{align}
	D_1&=-\frac{i}{2}\frac{\Gamma(\frac{1}{2})}{(4\pi)^{3/2}\Gamma(4)}\bigg(\frac{1}{|m_r|}\bigg)\Bigg[\frac{15}{4}-\frac{18}{4}+{\frac{3}{4}}\Bigg],
	{\notag}\\
	&=0.
\end{align}
\begin{figure}[ht!]	
	\includegraphics{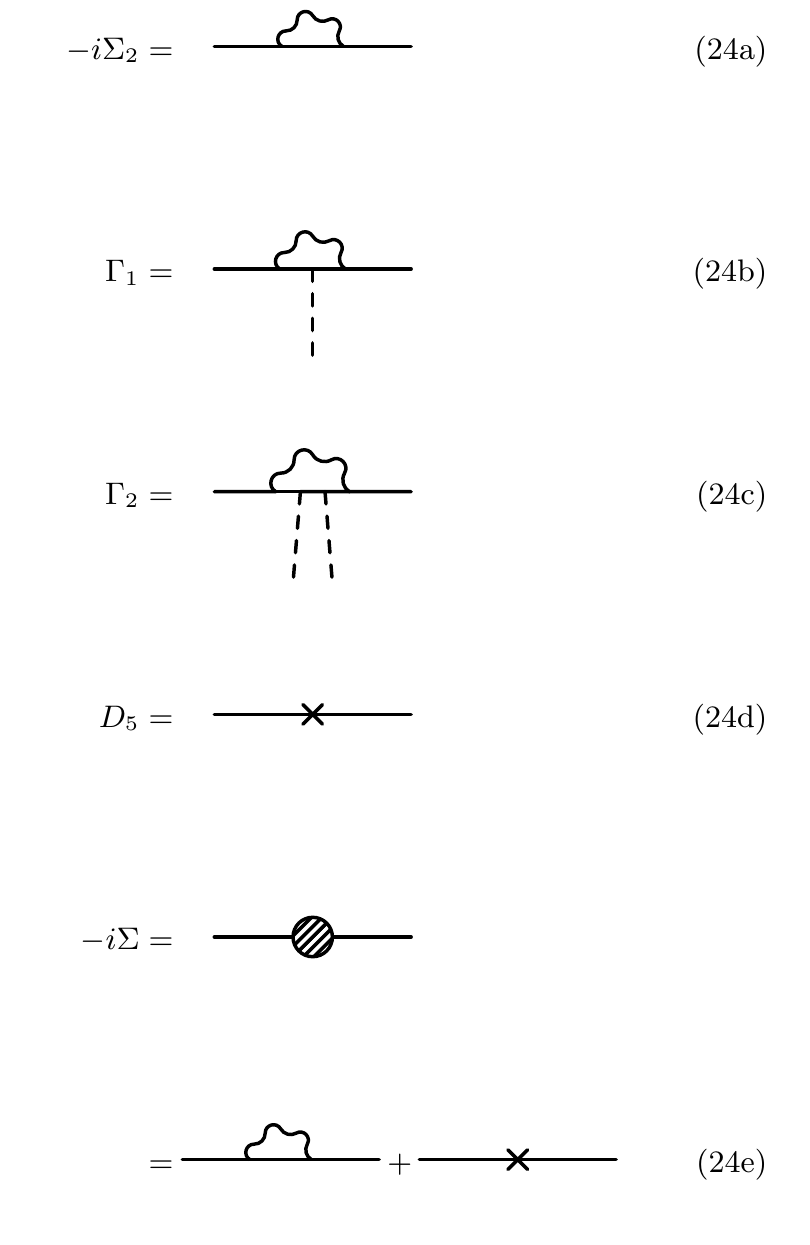}
	\caption{Electron self-energy and vertex corrections. Respectively, we have electron self-energy, $-i\Sigma_2$, the vertex correction $\Gamma_1$, $\Gamma_2$, and the counter terms  $D_5$. Those are also amputated.}\label{fig:els}
\end{figure}

As a preparation for the calculation of the two-loop diagrams, we first calculate the fermion self-energy, $\Sigma$, up to leading (one-loop) order in gauge coupling,
\begin{equation}
	-i\Sigma=-i\Sigma_2+D_5,\label{eq:selenr}
\end{equation}
where $-i\Sigma_2$ is shown in \cref{fig:els} and given as
\begin{align}
	-i\Sigma_2&=2\pi i\int \frac{d^3 q}{(2\pi)^3}\frac{\gamma_\mu(\slashed{q}-m_r)\gamma_{\nu}}{q^2-m_r^2+i\epsilon}\frac{\epsilon^{\mu\nu\alpha}(p-q)_\alpha}{(p-q)^2+i\epsilon},\label{eq:selfen}
\end{align}
and $D_5$ is the corresponding counter term.
 Equation (\ref{eq:selfen}) has a linear UV divergence. We can remove this by applying Pauli-Villars regularization, which is equivalent to the following substitution \cite{Peskin1995} in \cref{eq:selfen}:
\begin{equation}
	\frac{1}{(p-q)^2}\rightarrow \frac{1}{(p-q)^2}-\frac{1}{(p-q)^2-\Lambda^2},\label{eq:paulivil}
\end{equation}
where $\Lambda$ is the cutoff. Next, we use Feynman parameters to bring the denominator of  \cref{eq:selfen} in a spherically symmetric form by using the identity \cite{Peskin1995}
\begin{equation}
	\frac{1}{(q^2-m_{r}^2)^n(p-q)^2}=\int_{0}^{1}dx\frac{n(1-x)^{n-1}}{[(q-xp)^2-\Delta]^{n+1}},\label{eq:feynpam}
\end{equation}
where $\Delta=-p^2x(1-x)+(1-x)m_r^2$ and $n$ is a positive integer.  If we  substitute \cref{eq:paulivil} to \cref{eq:selfen}, apply \cref{eq:feynpam} with $n=1$ and change the integration variables as $q-xp\to k$, we have
\begin{align}
	&-i\Sigma_2= i\int_{0}^{1}dx\int \frac{d^3 k}{(2\pi)^2}\gamma_\mu(\slashed{k}+x\slashed{p}-m_r)\gamma_{\nu}\epsilon^{\mu\nu\alpha}{\notag}\\
	&\times\big(-k+p(1-x)\big)_\alpha\Bigg[\frac{1}{[k^2-\Delta]^2+i\epsilon}-\frac{1}{[k^2-\Delta_{\Lambda}]^2+i\epsilon}\Bigg],
\end{align}
where   $\Delta_{\Lambda}=-p^2x(1-x)+(1-x)m_r^2+\Lambda^2x$.
Next, we   use  the following identities for 2D gamma matrices:
\begin{subequations}
\begin{align}
	\epsilon^{\mu\nu\alpha}\gamma_\mu\slashed{a}\gamma_\nu&=2ia^\alpha,\\
	\epsilon^{\mu\nu\alpha}\gamma_\mu\gamma_\nu&=-2i\gamma^{\alpha}.
\end{align}\label{eq:gamid}
\end{subequations}
The self-energy is then given as
\begin{align}
	&-i\Sigma_2= -\int_{0}^{1}dx\int \frac{d^3 k}{2\pi^2}\Big(-k^2+(1-x)(p^2x+m_r\slashed{p})\Big){\notag}\\
	&\times\Bigg[\frac{1}{[k^2-\Delta]^2+i\epsilon}-\frac{1}{[k^2-\Delta_{\Lambda}]^2+i\epsilon}\Bigg],
\end{align}
where we have removed the terms that are odd in $k$. Next, we perform a Wick rotation and obtain
\begin{align}
	&-i\Sigma_2= -\frac{2i}{\pi}\int_{0}^{1}dx\int_{0}^{\infty} d k_E k_E^{2}\Big(k_E^2+(1-x){\notag}\\
	&\times(-p_E^2x-im_r\slashed{p}_E)\Big)\Bigg[\frac{1}{[k_E^2+\Delta_E]^2}-\frac{1}{[k_E^2+\Delta_{E\Lambda}]^2}\Bigg],
\end{align}
where $\Delta_E=p_E^2x(1-x)+(1-x)m_r^2$ and $\Delta_{E\Lambda}=p_E^2x(1-x)+(1-x)m_r^2+\Lambda^2x$. We can evaluate the integral over $k_E$ using
\begin{align}
I_1&=\int_{0}^{\infty} d k_E k_E^{4}\Bigg[\frac{1}{[k_E^2+\Delta_E]^2}-\frac{1}{[k_E^2+\Delta_{E\Lambda}]^2}\Bigg]{\notag},\\&=\frac{3\pi}{4}\Big(\sqrt{\Delta_{E\Lambda}}-\sqrt{\Delta_{E}}\Big),
\end{align}
and
\begin{align}
	I_2&=\int_{0}^{\infty} d k_E k_E^{2}\Bigg[\frac{1}{[k_E^2+\Delta_E]^2}-\frac{1}{[k_E^2+\Delta_{E\Lambda}]^2}\Bigg]{\notag},\\&=\frac{\pi}{4}\bigg(\frac{1}{\sqrt{\Delta_{E}}}-\frac{1}{\sqrt{\Delta_{E\Lambda}}}\bigg).
\end{align}
Note that  $\lim_{\Lambda\to\infty}\Delta_{E\Lambda}=\Lambda^2x$.
Finally, after  undoing the Wick rotation we have
\begin{align}
-i\Sigma_2&=-\frac{i}{2}\int_{0}^{1}dx\Bigg[-3\sqrt{\Delta}+\sqrt{x}\Lambda{\notag}\\&+\frac{1}{\sqrt{\Delta}}\Big(p^2x(1-x)+\slashed{p}m_{r}(1-x)\Big)\Bigg],
\end{align}
where we clearly see the linear divergence. The counter-terms will remove this divergence. We  define renormalization conditions as
\begin{subequations}
\begin{align}
	-i\Sigma(\slashed{p}=-m_r)&=0\\
	-i\frac{d\Sigma}{d\slashed{p}}\Big|_{\slashed{p}=-m_r}&=0,	
\end{align}\label{eq:rc}
\end{subequations}
which fixes the location of the poles and the residue, thus the physical mass \cite{Peskin1995}.
After substituting \cref{eq:selenr} to \cref{eq:rc}  we  have,
\begin{align}
	D_5&=i(\slashed{p}\delta_2+\delta m_r){\notag}\\
	&=i\Big(-\frac{1}{2}\text{sgn} (m_r)\slashed{p}+\frac{\Lambda}{3}-\frac{3}{2}|m_r|\Big). \label{eq:ccterm}
\end{align}

Next, we calculate $D_2$ which is shown in \cref{fig:phi4} and explicitly given as;
\begin{align}
	D_2&=-\frac{1}{4}\text{tr} \Bigg[\int \frac{d^3p}{(2\pi)^3}\frac{(i(\slashed{p}-m_r))^5}{(p^2-m_r^2)^5+i\epsilon}(-i\Sigma)\Bigg],\label{eq:fterm}
\end{align}
we then substitute \cref{eq:selenr,eq:selfen,eq:ccterm}, to \cref{eq:fterm}  perform a Wick rotation, and let $p_E\to p_E m_r$, which gives;
\begin{align}
	&D_2=\frac{i}{8\pi^2}\frac{1}{m_r}\int_{0}^{1}dx\int dp_{E}\  p^2_E\Bigg[(p_E^4-10p_E^2+5)(-p_E^2){\notag}\\ &\times\Big(1+\frac{(1-x)}{\sqrt{\Delta_{0}}}\Big)-(5p_E^4-10p_E^2+1){\notag}\\ &\times\Big(-3\sqrt{\Delta_0}+\frac{1-x}{\sqrt{\Delta}_0}(-p_E^2x)+3\Big)\Bigg]\frac{1}{(p_E^2+1)^5}, \label{eq:d2}
\end{align}
where $\Delta_0=(p_E^2x+1)(1-x)$.
It is easy to see that the sign of  $D_2$ depends on the combination of the sign of $m_r$ and sign of the level of the CS term, and the same is true for all two-loop contributions to $\beta_4$. Evaluating this integral yields
\begin{equation}
	D_2=\frac{i}{64\pi}\frac{1}{ m_r}.
\end{equation}

In preparation for the calculation of $D_3$, we first need to calculate the vertex correction $\Gamma_1$, which is shown in \cref{fig:els} and explicitly given as
\begin{align}
		\Gamma_{1}=&=2\pi \int \frac{d^3 q}{(2\pi)^3}\frac{\gamma_\mu(i(\slashed{q}-m_r))^2\gamma_{\nu}}{(q^2-m_r^2)^2+i\epsilon}\frac{\epsilon^{\mu\nu\alpha}(p-q)_\alpha}{(p-q)^2+i\epsilon}.\label{eq:g1}
\end{align}
Here, if we  check the superficial degree of divergence of $\Gamma_1$ by counting the net order of $q$, we naively find a logarithmic UV divergence. However, this is not the actual case, because the leading term of the integrand is an odd function of $q$. As a result, the naive logarithmically divergent term has  zero coefficient, and the integral in \cref{eq:g1} actually converges.
We apply \cref{eq:feynpam} with $n=2$, change the integration variables as $q-xp\to k$, and obtain
\begin{align}
\Gamma_1&=-4\pi\int_{0}^{1}dx\int \frac{d^3 k}{(2\pi)^3}\gamma_\mu(\slashed{k}+x\slashed{p}-m_r)^2\gamma_{\nu}\epsilon^{\mu\nu\alpha}{\notag}\\
&\times\big(-k+p(1-x)\big)_\alpha\Bigg[\frac{1-x}{[k^2-\Delta]^3+i\epsilon}\Bigg].
\end{align}
After using the Eqs. (\ref{eq:gamid}) and removing the odd terms in $k$, we have
\begin{align}
	\Gamma_1&=i8\pi\int_{0}^{1}dx\int \frac{d^3 k}{(2\pi)^3}\Bigg[\slashed{p}(1-x)(k^2+x^2p^2+m_r^2){\notag}\\
	&-\slashed{k}(2xp\cdot k)+2m_r(-k^2+(1-x)xp^2)\Bigg]{\notag}\\
	&\times\Bigg[\frac{1-x}{[k^2-\Delta]^3+i\epsilon}\Bigg].
\end{align} Now it is clear that this integral is not divergent, because the term that would produce logarithmic UV divergence is canceled as a result of the removal of the odd terms. We can further simplify this by  making the following substitution:
\begin{equation}
\slashed{k}(p\cdot k)\to\frac{1}{3}k^2\slashed{p},\label{eq:symint}
\end{equation}
which is a result of the symmetry of the integral in $k$. Then, we perform a Wick rotation:
\begin{align}
	&\Gamma_1=8\pi\int_{0}^{1}dx\int \frac{d^3 k_E}{(2\pi)^3}\Bigg[-i\slashed{p}_E(1-x)(-k_E^2-x^2p_E^2{\notag}\\
	&+m_r^2)-\frac{2k_E^2\slashed{p}_E}{3}+2m_r(k_E^2-(1-x)xp_E^2)\Bigg]\Bigg[\frac{1-x}{[k_E^2+\Delta_E]^3}\Bigg].
\end{align}
We can evaluate the integral over $k_E$ using the integral identities given in \cref{sec:inti}. We have,
\begin{align}
	\Gamma_1&=\frac{1}{4}\int_{0}^{1}dx(1-x)\bigg[\frac{3(i\slashed{p}_E(1-5x/3)+2m_{r})}{\sqrt{\Delta_E}}{\notag}\\ &+\frac{(m_r^2-x^2p^2_E)(1-x)(-i\slashed{p}_E)-2m_{r}x(1-x)p^2_E)}{\Delta_E^{3/2}} \bigg].
\end{align}
Finally, we can calculate  $D_3$, which is shown in \cref{fig:phi4} and explicitly given as
\begin{align}
	D_3&=-\frac{1}{4}\text{tr} \Bigg[\int \frac{d^3p}{(2\pi)^3}\frac{(i(\slashed{p}-m_r))^4}{(p^2-m_r^2)^4+i\epsilon}\Gamma_1\Bigg].
\end{align}
As before, we make the Wick rotation, let $p_E\to p_E m_r$, and substitute $\Gamma_1$, which gives
\begin{align}
	D_3&=-\frac{i}{16\pi^2}\frac{1}{m_r}\int_{0}^{1}dx\int_{0}^{\infty}dp_E p_E^2\Bigg[-p_E^2\Bigg(\frac{3-5x}{\sqrt{\Delta_0}}{\notag}\\&-\frac{(1-x)(1-x^2p_E^2)}{\Delta_0^{3/2}}\Bigg)(-4p_E^2+4){\notag}\\&+\Bigg(\frac{6}{\sqrt{\Delta_0}}-\frac{2(1-x)xp_E^2}{\Delta_0^{3/2}}\Bigg)(p_E^4-6p_E^2+1)\Bigg]{\notag}\\&\times\frac{1-x}{(p_E^2+1)^4} \label{eq:d3}.
\end{align}
Evaluating this integral yields
\begin{equation}
	D_3=0.
\end{equation}

Next we calculate $D_4$. First, we start with $\Gamma_2$, which is shown in \cref{fig:els} and explicitly given as
\begin{align}
	\Gamma_{2}=&=2\pi \int \frac{d^3 q}{(2\pi)^3}\frac{\gamma_\mu(i(\slashed{q}-m_r))^3\gamma_{\nu}}{(q^2-m_r^2)^3+i\epsilon}\frac{\epsilon^{\mu\nu\alpha}(p-q)_\alpha}{(p-q)^2+i\epsilon},
\end{align}
which is convergent.
First, we  apply \cref{eq:feynpam} with $n=3$ and change the integration variables as $q-xp\to k$, and we have
\begin{align}
	\Gamma_{2}=&=-i6\pi \int_{0}^{1}dx\int \frac{d^3 k}{(2\pi)^3}\gamma_\mu(\slashed{k}+x\slashed{p}-m_r)^3\gamma_{\nu}\epsilon^{\mu\nu\alpha}{\notag}\\
	&\times\big(-k+p(1-x)\big)_\alpha\Bigg[\frac{(1-x)^2}{[k^2-\Delta]^4+i\epsilon}\Bigg].
\end{align}
Next, we simplify the gamma matrix terms by using Eqs. (\ref{eq:gamid}) and show that
\begin{align}
&\epsilon^{\mu\nu\alpha}\gamma_\mu(\slashed{k}+x\slashed{p}-m_r)^3\gamma_{\nu}\big(-k+p(1-x)\big)_\alpha=2i(k^2{\notag}\\&+x^2p^2+2xk\cdot p+3m_r^2)(k+xp)\cdot(-k+p(1-x)){\notag}\\&+(-\slashed k+\slashed p(1-x))[3(k^2+x^2p^2+2xp\cdot k)+m_r^2]2im_r,\label{eq:gamsim}
\end{align}
then, we apply \cref{eq:symint} and a Wick rotation, so $\Gamma_2$ becomes
\begin{align}
	\Gamma_{2}&=12i\pi \int_{0}^{1}dx\int \frac{d^3 k_E}{(2\pi)^3}\Bigg(\Big(k_E^2-p_E^2x(1-x)\Big){\notag}\\
	&\times(-k_E^2-x^2p_E^2+3m_{r}^2)+\frac{2x}{3}p_E^2k_E^2(1-2x){\notag}\\
	&-im_r2xk^2\slashed{p}_E-im_r\slashed{p}_E(1-x)\Big(3(-k_E^2-x^2p_E^2)+m_r^2\Big)\Bigg){\notag}\\
	&\times\Bigg[\frac{(1-x)^2}{[k_E^2+\Delta_E]^4}\Bigg].
\end{align}
We can evaluate the integral over $k_E$ using  the integral identities given in \cref{sec:inti}.
$\Gamma_{2}$ is now
\begin{align}
	\Gamma_{2}&=\frac{3i}{16}\int_{0}^{1}dx(1-x)^2\Bigg[\frac{-5}{\Delta_{E}^{1/2}}+\frac{1}{\Delta_{E}^{3/2}}\bigg(\frac{5x}{3}p_E^2(1-2x){\notag}\\&+3m_r^2+im_r\slashed{p}_E(3-5x)\bigg)+\frac{1}{\Delta_{E}^{5/2}}\Bigg(-p_E^2x(1-x){\notag}\\&\times(-x^2p_E^2+3m_r^2)-im_r\slashed{p}_E(1-x)[-3x^2p_E^2+m_r^2]\Bigg)\Bigg].
\end{align}
Finally, we can calculate  $D_4$ which is shown in \cref{fig:phi4} and explicitly given as
\begin{align}
	D_4&=-\frac{1}{4}\text{tr} \Bigg[\int \frac{d^3p}{(2\pi)^3}\frac{(i(\slashed{p}-m_r))^3}{(p^2-m_r^2)^3+i\epsilon}(\Gamma_2)\Bigg].
\end{align}
As before we make Wick rotation, let $p_E\to p_E m_r$ and substitute $\Gamma_2$ which  gives
\begin{align}
	&D_4=-\frac{3i}{64\pi^2}\frac{1}{m_r}\int_{0}^{1}dx\int_{0}^{\infty}dp_E p_E^2\frac{(1-x)^2}{(p_E+1)^3}{\notag}\\
	&\times\Bigg\{-p_E^2(-p_E^2+3)\Bigg(\frac{3-5x}{\Delta_0^{3/2}}-(1-x)\frac{-3x^2p_E^2+1}{\Delta_0^{5/2}}+{\notag}\\
	&(1-3p_E^2)\Bigg[\frac{-5}{\Delta_{0}^{1/2}}+\frac{1}{\Delta_{0}^{3/2}}\bigg(\frac{5x}{3}p_E^2(1-2x){\notag}\\&+3\bigg)+\frac{1}{\Delta_{0}^{5/2}}\bigg(-p_E^2x(1-x)(-x^2p_E^2+3)\bigg)\Bigg]\Bigg)\Bigg\}.\label{eq:d4}
\end{align}
If we evaluate this numerically, we have
\begin{equation}
D_4=-\frac{i}{32\pi}\frac{1}{ m_r}.
\end{equation}

Finally, we  get $\beta_4$ by using \cref{eq:d2,eq:d3,eq:d4}  which  gives,
\begin{equation}
	\beta_4=-\frac{1}{64\pi}\frac{1}{ m_r},
\end{equation} which is the main result of this section.

We now discuss three different cases.

(i) $m_r > 0$. This describes a CSL phase. Since $\beta_4 > 0$, its transition into the phase with VBS and/or Ising Neel order is first order at the mean-field level.

(ii) $m_r < 0$. This describes an XY phase. Since $\beta_4 < 0$, its transition into the phase with VBS and/or Ising Neel order is second order at the mean-field level.

It should be noted that our evaluation of $\beta_4$ is only to the lowest order in gauge coupling (or inverse CS level), which is of order one. We cannot rule out the possibility that higher order correction can reverse the sign of $\beta_4$ and thus the conclusions above.

(iii) $m_r =0$. This is our multi-critical point, at which we can not integrate out the (massless) Dirac fermions perturbatively as done above. One can, nevertheless, perform a non-perturbative calculation of the effective potential \cite{Peskin1995} $V_{\rm eff}(\phi_{\rm cl})$ in terms of $\phi_{\rm cl}$, which is the vacuum expectation value of $\phi$ where $V_{\rm eff}(\phi_{\rm cl})$ is minimized. Since the fermion theory is massless and contains no scale, one expects its coupling to $\phi_{\rm cl}$ generates a scale-invariant term $|\phi_{\rm cl}|^3$, which is easy to verify by calculating the change of fermion ground-state energy due to $\phi_{\rm cl}$ that plays the role of a mass. The non-analyticity of such a term originates from the masslessness of the Dirac fermion. Its presence signals the non-mean-field behavior of the transition into the phases with broken translation symmetry, even if the theory is analyzed at the mean-field level.

\section{Dual Description}
\label{sec:Dual Description}
The theory of  multi-critical point is also discussed  in Ref. \cite{Wang2017a}, which is mainly done by considering a mean-field approach by considering the dual version of the theory. Thus, to have a connection with the literature, we also briefly find a dual version of our theory.
In Sec. \ref{sec:model}, we started with the lattice spin model given in \cref{eq:model1}, then we mapped it to hard-core bosons, and then mapped those hard-core bosons to  non-relativistic  fermions in a lattice with  a level-one CS term. Then, we found that the continuum limit of  this theory is described by two Dirac fermions coupled to the level-one CS term given in \cref{eq:Low-energy theory}.
In this section,  we will apply a bosonization  transformation to \cref{eq:Low-energy theory}, which, in a sense, close the circle of our mappings.

We will use the well-known bosonization conjecture   \cite{Karch2016,DualityReview,tong2016d,Seiberg,Son2019,Raghu2018b}. First we have to make several definitions to simplify the notation in the following calculations. We closely follow the approach of Ref. \cite{Karch2016} in this section.
We define the CS term and background field coupling as \cite{Karch2016}
\begin{subequations}
	\begin{align}
		S_{\text{CS}}[a]&=\frac{1}{4\pi}\int d^3x\epsilon^{\mu\nu\lambda}a_\mu \partial_\nu a_\lambda,\\
		S_{\text{BF}}[a,B]&=\frac{1}{2\pi}\int d^3x\epsilon^{\mu\nu\lambda}a_\mu \partial_\nu B_\lambda, \label{eq:bf},
	\end{align}
\end{subequations}
where $a$ is a dynamic gauge field and $B$ is a background gauge field  note that we use  lower case letters for dynamic gauge fields and upper case  letters for background-gauge fields as before. The actions for material fields are given as
\begin{subequations}
	\begin{align}
		S_{\text{fermion}}[\psi,A]&=\int d^3x\bar{\psi}(i(\gamma^\mu\partial_\mu-iA_\mu)\psi,\\
		S_{\text{scalar}}[\phi,A]&=\int d^3x|(\partial_\mu-iA_\mu)\phi|^2-\alpha|\phi|^4, \label{eq:scalaraction}
	\end{align}
\end{subequations}
where we have an action for a free Dirac fermion coupled to the background gauge field and complex Wilson-Fischer (WF) scalar, with coupling constant $\alpha$  which flows to infinity at the WF fixed point and the mass flows to zero \cite{Karch2016}. Their partition functions
\begin{subequations}
	\begin{align}
		Z_{\text{fermion}}[A]&=\int \mathcal{D}\bar{\psi}\mathcal{D}\psi e^{iS_{\text{fermion}}[\psi,A]},\\
		Z_{\text{scalar}}[A]&=\int \mathcal{D}\bar{\phi}\mathcal{D}\phi e^{iS_{\text{scalar}}[\phi,A]},
	\end{align}
\end{subequations}
are related by the bosonization conjecture \cite{Karch2016}:
\begin{equation}
	Z_{\text{fermion}}[A]e^{-\frac{i}{2}S_{\text{CS}}[A]}=\int \mathcal{D}a Z_{\text{scalar}}[a]e^{iS_{\text{CS}}[a]+iS_{\text{BF}}[a;A]}.
	\label{eq:3dbos}
\end{equation}
We have to clarify the origin of the extra half-level CS term on the left-hand-side (LHS), which is something purely notational. To understand this, assume for a moment the fermions in LHS are massive. In our  notation when we integrate out the fermions of $	Z_{\text{fermion}}[A]$, we do not perform any Pauli-Villars regularization \cite{Tong2018}, and  as a result of that, we get  a half-level CS term after  integrating out the fermions. However, a CS term with a noninteger level breaks the large gauge invariance \cite{Tong2018}. So, to preserve the gauge invariance of the theory, we have to add that extra half-level CS term \footnote{We could have used an alternative notation such that when we integrate out the fermions in $Z_f[A]$ we could have performed Pauli-Villars regularization. In that case, Pauli-Villars regularization would automatically add that extra half-level CS term which  would eliminate to necessity of adding it by hand. However, in his paper we don't use this notation.}.

Our goal is to obtain \cref{eq:Low-energy theory} by applying a series of manipulations to LHS of \cref{eq:3dbos}. Applying the same manipulations to the right-hand-side (RHS) of \cref{eq:3dbos} yields the dual (or bosonized) version of \cref{eq:Low-energy theory}.

First, we multiply two copies of the LHS of  \cref{eq:3dbos} and integrate it over $A$. We denote this integration variable as  $\tilde{a}$, and  add a coupling with background field $C$. So, the theory becomes,
\begin{equation}
	S_L=	S_{\text{f}}[\psi_1,\tilde{a}]+S_{\text{f}}[\psi_2,\tilde{a}]-S_{\text{BF}}[\tilde{a},C]-S_{\text{CS}}[\tilde{a}],
\end{equation}
which gives the fermionic side of our new duality. Performing the same manipulations to the RHS of \cref{eq:3dbos} yields
\begin{align}
	S_R=&	S_{\text{s}}[\phi_1,a_1]+S_{\text{s}}[\phi_2,a_2]+	S_{\text{CS}}[a_1]+	S_{\text{CS}}[a_2]{\notag}\\&+S_{\text{BF}}[a_1+a_2-C,\tilde{a}],
\end{align}
and this is the bosonic side of our new duality. Next, we integrate out $\tilde{a}$ on the RHS, which gives rise to the constraint
\begin{equation}
	C=a_1+a_2,
\end{equation}
which we solve by introducing a new dynamic field $b$ as $a_1=b$ and $a_2=-b+C$.
Then $S_R$ becomes
\begin{equation}
	S_R=	S_{\text{s}}[\phi_1,b+	]+S_{\text{s}}[\phi_2,-b+C]+	S_{\text{CS}}[b]+	S_{\text{CS}}[-b+C].
\end{equation}
Next, we apply time reversal transformation to both sides by simply changing the signs of the BF and CS terms.  We then have,
\begin{equation}
	S_L'=	S_{\text{f}}[\psi_1,\tilde{a}]+S_{\text{f}}[\psi_2,\tilde{a}]+S_{\text{BF}}[\tilde{a},C]+S_{\text{CS}}[\tilde{a}].
\end{equation}
The motivation behind this transformation is clear, as the fermionic theory now contains a level-one CS term as in \cref{eq:Low-energy theory}. Accordingly, the bosonic  side of the duality is,
\begin{equation}
	S_R'=	S_{\text{s}}[\phi_1,b]+S_{\text{s}}[\phi_2,-b+C]-	S_{\text{CS}}[b]-	S_{\text{CS}}[-b+C].
\end{equation}
Next, we let $C\to-C$ and we add $S_{\text{CS}}[C]$ to  both sides. So, both sides of the duality are given as,
\begin{equation}
	S_L''=	S_{\text{f}}[\psi_1,\tilde{a}]+S_{\text{f}}[\psi_2,\tilde{a}]+S_{\text{CS}}[\tilde{a}-C].
\end{equation}
Note that for $C=A$ this is just the action of \cref{eq:Low-energy theory} and the bosonic side of the duality is,
\begin{align}
	S_R''&=	S_{\text{s}}[\phi_1,b]+S_{\text{s}}[\phi_2,-b-C]{\notag}\\&-2S_{\text{CS}}[b]-S_{\text{BF}}[b,C].
\end{align}
Finally, we let $\phi_2 \leftrightarrow \phi^{*}_2$ and get,
\begin{align}
	S_R'''&=	S_{\text{s}}[\phi_1,b]+S_{\text{s}}[\phi_2,b+C]{\notag}\\&-2S_{\text{CS}}[b]-S_{\text{BF}}[b,C],
\end{align}
which concludes the bosonization of \cref{eq:Low-energy theory}. One should note that this is not the only possible duality that one can find. For example, we can find different bosonic dual models to our original model by considering the time-reversed  version of \cref{eq:3dbos} to the one of the fermionic degrees of freedom in our original model.

\section{Summary and Discussion}
\label{sec:Summary}

In this paper we provide a unified description of various possible phases supported by a spin-$\frac{1}{2}$ antiferromagnet with easy-plane anisotropy on the square lattice, including Neel-order states, CSL, and VBSs. The description is based on two Dirac fermions coupled to a level-one CS gauge field, and the various phases correspond to different combinations of the various Dirac mass terms. All these phases meet at a multi-critical point where the entire Dirac mass matrix vanishes. Within our description, a direct continuous transition from the XY-ordered Neel state to the VBS must go through this multi-critical point. In more generic situations there is either an intermediate phase with both orders, or a direct first-order transition.

The theory of this multi-critical point and its dual description have some similarities to that of the deconfined criticality\cite{deconfined_cirticalityPRB} and its dual description\cite{Wang2017a}. The main difference is our models contain CS couplings, while their models do not. As a result, their phase diagram does not contain the CSL phase.

\subsection*{Note Added}

After this manuscript is published, we became aware of three closely related papers of Wang \emph{et. al.} \cite{wang1,wang2,wang3}, discussing unified explanation of Neel-AFM state, CSL and corresponding phase transitions in the mean-field theory of fermions coupled with CS theory and diligent studies of the ordered states. 

\section*{Acknowledgments}
This work was initiated at Stanford University during the one of K.Y.'s sabbatical leave there, and he thanks Profs. Steve Kivelson, Sri Raghu and especially late Shoucheng Zhang for their invitation and hospitality, as well as Stanford Institute of Theoretical Physics and Gordon and Betty Moore Foundation for support. He also benefited from stimulating discussions with Jingyuan Chen, Jun-Ho Son, Sri Raghu, and T. Senthil.
This work was supported by the National Science Foundation Grant No. DMR-1932796, and performed at the National High Magnetic Field Laboratory, which is supported by National Science Foundation Cooperative Agreement No. DMR-1644779, and the State of Florida.

\section*{APPENDIX: Integral identities}\label{sec:inti}
Here we discuss the common integrals we will encounter in the main text \cite{Peskin1995},
\begin{widetext}
	\begin{subequations}
	\begin{align}
	\int \frac{d^d k_E}{(2\pi)^d} \frac{1}{[k_E^2+\Delta_E]^n}&=\frac{1}{(4\pi)^{d/2}}\frac{\Gamma(n-\frac{d}{2})}{\Gamma(n)}\bigg(\frac{1}{\Delta_{E}}\bigg)^{n-\frac{d}{2}},\\
	\int \frac{d^d k_E}{(2\pi)^d} \frac{k_E^2}{[k_E^2+\Delta_E]^n}&=\frac{d}{2(4\pi)^{d/2}}\frac{\Gamma(n-\frac{d}{2}-1)}{\Gamma(n)}\bigg(\frac{1}{\Delta_{E}}\bigg)^{n-\frac{d+2}{2}},\\
	\int \frac{d^d k_E}{(2\pi)^d} \frac{k_E^{4}}{[k_E^2+\Delta_E]^n}&=\frac{d(d/2+1)}{2(4\pi)^{d/2}}\frac{\Gamma(n-\frac{d}{2}-2)}{\Gamma(n)}\bigg(\frac{1}{\Delta_{E}}\bigg)^{n-\frac{d+4}{2}},
\end{align}
\end{subequations}
\end{widetext}
 where $n\in\mathbb{Z}^+$, which can be proved easily by converting the LHS to the Euler integral (Beta function) by substituting $x=\Delta_E/(k_E^2+ \Delta_E)$.
\

\end{document}